\pdfoutput=1  

\def\virg#1{\textquoteleft #1\textquoteright}

\def\tn{\textnormal{-}}

\def\acap{\\ \nonumber \\}

\def\rfr#1{Equation\,(\ref{#1})}
\def\rfrs#1#2{Equations\,(\ref{#1})--(\ref{#2})}

\def\dert#1#2{\frac{{{\mathit{d}}}{#1}}{{{\mathit{d}}}{#2}}} 
 
\def\eqi{\begin{equation}}
\def\eqf{\end{equation}}
\def\eqia{\begin{eqnarray}}
\def\eqfa{\end{eqnarray}}
\def\rp#1#2{\frac{#1}{#2}}
\def\lb#1{\label{#1}}

\def\nk{n_\mathrm{K}}

\def\ton#1{\left(#1\right)}
\def\qua#1{\left[#1\right]}
\def\grf#1{\left\{#1\right\}}

\documentclass[a4paper, 11pt]{article}

\usepackage[english]{babel}
\usepackage[LGR,T1]{fontenc}
\usepackage[utf8]{inputenc}
\usepackage{w-greek}
\usepackage[nointegrals]{wasysym}
\usepackage{authblk}
\usepackage{abstract}
\usepackage{amsmath}
\usepackage{amsthm}
\usepackage{amscd}
\usepackage{amssymb}
\usepackage{dsfont}
\usepackage{xcolor}
\usepackage{graphicx}
\usepackage{epsfig}
\usepackage{xr-hyper}
\usepackage[pdfstartview=XYZ,
bookmarks=true,
colorlinks=true,
linkcolor=blue,
urlcolor=blue,
citecolor=blue,
pdftex,
bookmarks=true,
linktocpage=true, 
hyperindex=true
]{hyperref}
\usepackage{starfont}
\usepackage{txfonts}
\usepackage[mathlines]{lineno}
\usepackage{tabularx}
\usepackage{booktabs}
\usepackage{fontawesome5}

\usepackage[round]{natbib} 
\allowdisplaybreaks[1]

\makeatletter
\DeclareRobustCommand\ref{%
    \@ifstar\@refstar\T@ref
  }%
  \DeclareRobustCommand\pageref{%
    \@ifstar\@pagerefstar\T@pageref
  }%
 \makeatother

\DeclareSymbolFont{greekletters}{LGR}{\familydefault}{m}{n}
\DeclareMathSymbol{\qA}{\mathord}{greekletters}{65}
\DeclareMathSymbol{\qB}{\mathord}{greekletters}{66}
\DeclareMathSymbol{\qG}{\mathord}{greekletters}{71}
\DeclareMathSymbol{\qD}{\mathord}{greekletters}{68}
\DeclareMathSymbol{\qE}{\mathord}{greekletters}{69}
\DeclareMathSymbol{\qZ}{\mathord}{greekletters}{90}
\DeclareMathSymbol{\qEt}{\mathord}{greekletters}{72}
\DeclareMathSymbol{\qTh}{\mathord}{greekletters}{74}
\DeclareMathSymbol{\qI}{\mathord}{greekletters}{73}
\DeclareMathSymbol{\qK}{\mathord}{greekletters}{75}
\DeclareMathSymbol{\qL}{\mathord}{greekletters}{76}
\DeclareMathSymbol{\qM}{\mathord}{greekletters}{77}
\DeclareMathSymbol{\qN}{\mathord}{greekletters}{78}
\DeclareMathSymbol{\qX}{\mathord}{greekletters}{88}
\DeclareMathSymbol{\qO}{\mathord}{greekletters}{79}
\DeclareMathSymbol{\qP}{\mathord}{greekletters}{80}
\DeclareMathSymbol{\qR}{\mathord}{greekletters}{82}
\DeclareMathSymbol{\qS}{\mathord}{greekletters}{83}
\DeclareMathSymbol{\qT}{\mathord}{greekletters}{84}
\DeclareMathSymbol{\qU}{\mathord}{greekletters}{85}
\DeclareMathSymbol{\qPh}{\mathord}{greekletters}{70}
\DeclareMathSymbol{\qCh}{\mathord}{greekletters}{81}
\DeclareMathSymbol{\qPs}{\mathord}{greekletters}{89}
\DeclareMathSymbol{\qOm}{\mathord}{greekletters}{87}
\DeclareMathSymbol{\qa}{\mathord}{greekletters}{97}
\DeclareMathSymbol{\qb}{\mathord}{greekletters}{98}
\DeclareMathSymbol{\qg}{\mathord}{greekletters}{103}
\DeclareMathSymbol{\qd}{\mathord}{greekletters}{100}
\DeclareMathSymbol{\qe}{\mathord}{greekletters}{101}
\DeclareMathSymbol{\qz}{\mathord}{greekletters}{122}
\DeclareMathSymbol{\qet}{\mathord}{greekletters}{104}
\DeclareMathSymbol{\qth}{\mathord}{greekletters}{106}
\DeclareMathSymbol{\qi}{\mathord}{greekletters}{105}
\DeclareMathSymbol{\qk}{\mathord}{greekletters}{107}
\DeclareMathSymbol{\ql}{\mathord}{greekletters}{108}
\DeclareMathSymbol{\qm}{\mathord}{greekletters}{109}
\DeclareMathSymbol{\qn}{\mathord}{greekletters}{110}
\DeclareMathSymbol{\qx}{\mathord}{greekletters}{120}
\DeclareMathSymbol{\qo}{\mathord}{greekletters}{111}
\DeclareMathSymbol{\qp}{\mathord}{greekletters}{112}
\DeclareMathSymbol{\qr}{\mathord}{greekletters}{114}
\DeclareMathSymbol{\fs}{\mathord}{greekletters}{99}       
\DeclareMathSymbol{\qs}{\mathord}{greekletters}{115}       
\DeclareMathSymbol{\qt}{\mathord}{greekletters}{116}
\DeclareMathSymbol{\qu}{\mathord}{greekletters}{117}
\DeclareMathSymbol{\qvf}{\mathord}{greekletters}{102}
\DeclareMathSymbol{\qch}{\mathord}{greekletters}{113}
\DeclareMathSymbol{\qps}{\mathord}{greekletters}{121}
\DeclareMathSymbol{\qom}{\mathord}{greekletters}{119}
\DeclareMathSymbol{\sui}{\mathord}{greekletters}{124}
\DeclareMathSymbol{\dig}{\mathord}{greekletters}{147}
\DeclareMathAccent{\uml}{\mathord}{greekletters}{34} 
\DeclareMathAccent{\umld}{\mathord}{greekletters}{35} 
\DeclareMathAccent{\mcr}{\mathord}{greekletters}{31}
\DeclareMathAccent{\aca}{\mathord}{greekletters}{39} 
\DeclareMathAccent{\ga}{\mathord}{greekletters}{96} 
\DeclareMathAccent{\rb}{\mathord}{greekletters}{60} 
\DeclareMathAccent{\smb}{\mathord}{greekletters}{62} 
\DeclareMathAccent{\ca}{\mathord}{greekletters}{126} 
\DeclareMathAccent{\carb}{\mathord}{greekletters}{64} 
\DeclareMathAccent{\casb}{\mathord}{greekletters}{92} 
\DeclareMathAccent{\aarb}{\mathord}{greekletters}{86} 
\DeclareMathAccent{\aasb}{\mathord}{greekletters}{94} 
\DeclareMathAccent{\garb}{\mathord}{greekletters}{67} 
\DeclareMathAccent{\gasb}{\mathord}{greekletters}{95} 

\DeclareTextSymbol{\textquoteleft}{LGR}{28}      
\DeclareTextSymbol{\textquoteright}{LGR}{29}     

\definecolor{orcidlogocol}{rgb}{0.65, 0.807, 0.223}

\newcommand{\orcid}[1]{$\,$\href{https://orcid.org/#1}{\textcolor{orcidlogocol}{\faOrcid}}}

\title{
\textbf{
The no-hair theorems at work in the tidal disruption event AT2020afhd}
}

\author[]{
Lorenzo Iorio\,\orcid{0000-0003-4949-2694}
}
\affil[]{
\href{https://ror.org/01ehyh486}{Ministero dell' Istruzione e del Merito}
\\
Viale Unit\`{a} di Italia 68, I-70125, Bari, Italy \\ email: \href{mailto:lorenzo.iorio@libero.it}{\texttt{lorenzo.iorio@libero.it}}
}

\date{\today}

\providecommand{\keywords}[1]{keywords--- #1}

\begin{document}

\maketitle

\begin{center}
\begin{abstract}
\noindent
Recently, the coprecession of both the accretion disk and the jet formed following the tidal disruption event associated with the optical transient  AT2020afhd, driven by a  supermassive black hole of almost ten million solar masses,  were independently measured in both the X and radio bands, respectively, showing a periodicity of nearly 20 days over about 300 days. An analytical model of the general relativistic gravitomagnetic Lense-Thirring precession of the effective orbit of a fictitious test particle revolving about a spinning primary can explain the observed precessional features. It yields allowed regions in the system's parameter space which, as far as the hole's dimensionless spin parameter is concerned, are essentially in agreement with those obtained in the literature with general relativistic magnetohydrodynamic simulations. The present analytical approach can be extended to include  the precession due to the hole's quadrupole mass moment as well. It breaks the degeneracy in the allowed regions occurring for negative and positive values of the spin parameter when only the Lense-Thirring effect is considered. The best estimate for the hole's mass yields the range $0.185-0.215$ for the dimensionless spin parameter.
Using the same strategy with the gravitomagnetic frequency for an extended disk of finite size with a parameterized power-law mass density yields to distinct, generally non-overlapping allowed regions for each value of the power-law index adopted. \textcolor{black}{Some of the assumptions on which this work is based are critically examined.}
\end{abstract}
\end{center}

\section{Introduction}
Tidal disruption events (TDEs), first theoretically predicted by \citet{1975Natur.254..295H}, occur when main sequence stars, passing by a supermassive black hole (SMBH), or, more compactly, megapyknon (MP) \citep{2025Univ...11..251I}, like those that lurk in the core of several galaxies, are torn apart by the tidal forces due to the intense gradient of the gravitational field \citep{1988Natur.333..523R,1989ApJ...346L..13E,tde2026}. The resulting debris arranges itself in orbit around the black hole (BH), forming temporarily  a rapidly circularized \citep{2016MNRAS.461.3760H} thick accretion disk \citep{1999ApJ...514..180U,2009MNRAS.400.2070S},  initially inclined with respect to the hole's equatorial plane \citep{1990ApJ...351...38C}, around the tidal radius. Over time, it tends to align with the hole's spin due to concomitant magnetic forces acting on the ionized particles of the disk itself \citep{2013Sci...339...49M}. The resulting complex physical environment around the accreting BH \citep{2021SSRv..217...12D} can also sometimes induce the formation of a moderately relativistic jet \citep{2020NewAR..8901538D,2022FrASS...8..234F,2024Univ...10..156F} that, for periods ranging from a few months to a few years, is often visible by emitting electromagnetic waves in the radio band, while the accreting matter pour outs X-rays. As long as the disk's misalignment with respect to the equator persists, complex  general relativistic magnetohydrodynamic (GRMHD) simulations predict that the jet and the disk remain tightly coupled \citep{2018MNRAS.474L..81L,2020MNRAS.499..362C}, exhibiting a coprecessional motion that is commonly attributed to the general relativistic gravitomagnetic Lense-Thirring (LT) effect \citep{2007ApJ...668..417F,2012PhRvL.108f1302S,2016MNRAS.455.1946F,2024Natur.630..325P}. 
At this point, it may be useful to clarify that denominations like \virg{gravitomagnetism} have nothing to do with any interplay between gravitation and genuine magnetic forces, being just due to the formal resemblance of the linearized Einstein field equations of general relativity in the weak-field and slow-motion limit with the linear Maxwell's ones for electromagnetism \citep{Thorne86,1986hmac.book..103T,1988nznf.conf..573T,1992AnPhy.215....1J,2001rfg..conf..121M,2002NCimB.117..743R,Mash07}. Within such an approximation, the off-diagonal terms of the spacetime metric tensor give rise to an acceleration felt by a test particle orbiting a massive rotating body which looks like that due to the Lorentz force experienced by an electric charge moving in a magnetic field.
Within the context of TDEs, the LT effect leaves a characteristic signature in both the X-ray and radio emission in the form of  quasi-periodic modulations that, up to now, have never been observed simultaneously, having instead been measured either in X-rays \citep{2012MNRAS.422.1625S,2021NatAs...5...94M,2024Natur.630..325P} or in the radio band \citep{2023Natur.621..711C,2023Natur.621..271T} separately. 

This situation has recently changed with the TDE associated with the substantial rebrightening\footnote{Its observed optical decay rate is in agreement with the debris fallback rate expected in TDE theories.} in 2024 of the optical transient AT2020afhd/ZTF20abwtifz that occurred in 2020 in the galaxy LEDA 145386 where both X-ray and radio signals from the disk-jet coprecession were simultaneously measured for the first time \citep{TDELTCinesi2025}.
The resulting picture, emerging from a data analysis spanning about 300 days,  encompasses a Kerr SMBH of about ten million solar masses which, rotating rather slowly 
dragged a mildly misaligned  disk into a precession with a period of about 20 days; see Table \ref{Tab:1} for the relevant parameters of this TDE.
\begin{table}[ht]
\centering
\caption{Physical and orbital parameters of the TDE associated with AT2020afhd as determined by \citet{TDELTCinesi2025}. The mass of the SMBH is $M_\bullet$, while the Sun's mass is denoted by $M_\odot$. The period of the disk-jet coprecession, in days, is $T_\mathrm{prec}$. The angle $\theta_i$ is counted from the reference $z$ axis of the coordinate system adopted in \textbf{Fig. 5} (\textbf{A}) of \citet{TDELTCinesi2025}, whose direction is assumed equal to that of the SMBH'symmetry axis, to the unit vector $\boldsymbol{\hat{h}}$ of the disk's orbital angular momentum, while $\phi_\mathrm{obs},\theta_\mathrm{obs}$ are the longitude and colatitude, respectively, of the line of sight.
}\lb{Tab:1}
\vspace{0.3cm}
\begin{tabular}{l l l l l}
\toprule
$\log\ton{M_\bullet/M_\odot}$ & $T_\mathrm{prec}$ (d) & $\theta_i$ ($^\circ$) & $\phi_\mathrm{obs}$ ($^\circ$) & $\theta_\mathrm{obs}$ ($^\circ$)\\
\midrule
$6.7\pm 0.5$ & $19.6\pm 1.5$ & $14.5\pm 0.5$ & $188.4\pm 1.5$ & $38.4_{-0.6}^{+0.5}$  \\
\bottomrule
\end{tabular}
\end{table}

In this paper, it is shown that it is possible to obtain the same results about the LT hypothesis and the SMBH's spin with a very simple analytical model, formally accurate to the first post Newtonian (1pN) level, of the gravitomagnetic precession of the orbital angular momentum of a fictitious test particle moving along a sort of \virg{effective} circular orbit \citep{1974PhRvD..10.1340B,2025PhRvD.111d4035I} whose radius equals the tidal disruption one for the SMBH considered. Such a picture could even be something more than a simple effective phenomenological artifice since GRMHD simulations show that the disks expected in TDEs do, indeed, rotate rigidly \citep{2007ApJ...668..417F,2011ApJ...730...36D}. It can be further enriched by quantifying also the contribution of the SMBH's quadrupole mass moment \citep{2025MNRAS.537.1470I} in the present case, neglected by \citet{TDELTCinesi2025}. The same approach, which can be straightforwardly extended to other systems like the one considered here, was successfully adopted in \citet{2025MNRAS.537.1470I,2025PhRvD.111d4035I} also for the recently measured jet precession emanating from the surrounding of the SMBH M87$^\ast$ \citep{2023Natur.621..711C,2025NatAs...9.1218C}. \textcolor{black}{If confirmed by further analyses,} it may be viewed as another unexpected example of the \virg{the unreasonable effectiveness of the post-Newtonian approximation in gravitational physics} \citep{2011PNAS..108.5938W}. \textcolor{black}{
Indeed, it must be recognized that, in principle, other physical mechanisms exist besides the Lense-Thirring effect that could potentially explain the observed phenomenon. These include, for example, disk oscillations, orbiting hot spots, and jet instabilities. The magnetic field that is usually present around SMBHs can also be a cause of precession for the disk composed of ionized matter. Finally, another massive companion at the right distance could also induce disk precession. Examining in detail all these possible alternative causes other than the Lense-Thirring effect alone is beyond the scope of this article, as they deserve one or more dedicated works.
}

The paper is organized as follows. The mathematical model assumed for effectively describing the disk-jet coprecession is described in Section \ref{Sec:2}. It is applied to infer constraints on the system's parameters in Section \ref{Sec:3}. Section \ref{Sec:4} is devoted to an analysis of the bounds obtained by \citet{TDELTCinesi2025} with a disk of finite size endowed with a power-law density profile. \textcolor{black}{In Section \ref{Sec:5}, some of the assumptions on which this work mainly relies upon are critically scrutinized.}
Section\,\ref{Sec:6} summarizes the present findings and offers conclusions.
\section{The mathematical model of the no-hair disk precession}\lb{Sec:2}
The precession of the unit vector $\boldsymbol{\hat{h}}$ of the orbital angular momentum of a test particle orbiting a massive object of mass $M$, angular momentum $J$ and dimensional quadrupole mass moment $Q_2$, with $\qua{Q_2} = \mathrm{M}\mathrm{L}^2$, is
\eqi
\dert{\boldsymbol{\hat{h}}}{t} = {\boldsymbol{\Omega}}\boldsymbol\times\boldsymbol{\hat{h}},\lb{dhdt}
\eqf
In it, the angular velocity vector $\boldsymbol{\Omega}$ of the precession is
\eqi
{\boldsymbol{\Omega}} = \Omega\boldsymbol{\hat{k}},
\eqf
where $\boldsymbol{\hat{k}}$ is the primary's spin axis, and \citep{2025MNRAS.537.1470I,2025PhRvD.111d4035I}
\eqi
\Omega = \Omega_\mathrm{LT} + \Omega_{Q_2} = \rp{2GJ}{c^2 p^3\ton{1-e^2}^{-3/2}} +\rp{3\nk Q_2}{2 M p^2}\boldsymbol{\hat{k}}\boldsymbol\cdot\boldsymbol{\hat{h}}.\lb{ONH}
\eqf
In \rfr{ONH}, $G$ is the Newtonian constant of gravitation, $c$ is the speed of light in vacuum, $e$ is the orbital eccentricity, $p$ is the orbital semilatus rectum which, in the limit $e\rightarrow 0$, reduces to the orbital radius $r_0$, and 
\eqi
\nk:=\sqrt{\rp{GM\ton{1-e^2}^3}{p^3}}
\eqf
is the Keplerian mean motion which can be rewritten as 
\eqi
\nk=\sqrt{\rp{GM}{r_0^3}}
\eqf
for a circular orbit. 

In the case of the Kerr metric \citep{1963PhRvL..11..237K,1965qssg.conf...99K,Visser09,2015CQGra..32l4006T}, which is believed to describe the exterior spacetime of a spinning BH \citep{1970Natur.226...64B}, the so-called \virg{no-hair} (NH) theorems \citep{1967PhRv..164.1776I,1971PhRvL..26..331C,1975PhRvL..34..905R}, as per the fortunate aphorism by J.A. Wheeler \citep{2009PhT....62d..47R}, predict that, contrary to an ordinary nonspherical material body, all the mass and spin multipole moments $\mathcal{M}_\bullet^\ell$ and $\mathcal{J}_\bullet^\ell$  of degree $\ell=0,1,2,\ldots$, respectively,  can be expressed as\footnote{The formalisms by Geroch-Hansen and Thorne were proven to be equivalent, up to a normalization factor, by \citet{1983GReGr..15..737G}.} \citep{1970JMP....11.2580G,1974JMP....15...46H,1980RvMP...52..299T}
\eqi
\mathcal{M}_\bullet^\ell + i\mathcal{J}_\bullet^\ell = M_\bullet\ton{i\rp{J_\bullet}{cM_\bullet}}^\ell,\lb{MQ}
\eqf
where $i:=\sqrt{-1}$ is the imaginary unit. From \rfr{MQ}, it turns out that the odd mass moments and the even spin moments vanish; the mass moment of degree $\ell=0$ is nothing but the BH's mass, while the spin moment of degree $\ell=1$ is proportional to its spin angular momentum.
For a Kerr BH, the latter is \citep{1986bhwd.book.....S}
\eqi
J_\bullet  = a_\bullet \rp{M_\bullet^2 G}{c}; \lb{SBH}
\eqf
\rfr{SBH} comes from the condition that the $g_{rr}$ component of the Kerr metric in the Boyer-Lindquist coordinates \citep{1967JMP.....8..265B} goes to infinity.
The quadrupole mass moment, namely the mass moment $\mathcal{M}_\bullet^2$ of degree $\ell=2$ in \rfr{MQ} renamed here as $Q_2^\bullet$, is 
\eqi
Q_2^\bullet  = -\rp{J^2_\bullet}{c^2 M_\bullet}.\lb{QBH}
\eqf 
In \rfr{SBH}, $a_\bullet$ is the dimensionless spin parameter \citep{1972ApJ...178..347B} which can be viewed as the second characteristic length\footnote{Dubbed as angular momentum per unit mass or specific angular momentum when geometrized units $G=c=1$ are adopted, $l_\mathrm{g}$ is sometimes denoted in the literature by the same symbol $a_\bullet$ which, instead, is used in this paper for its dimensionless counterpart. Caution is in order to avoid confusions with the semimajor axis of a Keplerian orbit, usually denoted as $a$ in celestial mechanics \citep{2000ssd..book.....M}.} 
\eqi
l_\mathrm{g}:=\rp{J_\bullet}{cM_\bullet}
\eqf
entering the Kerr metric measured in units of the BH's gravitational radius 
\eqi
R_\mathrm{g}:=\rp{GM_\bullet}{c^2}.
\eqf
The value of $a_\bullet$ must fulfil the condition
\eqi
\left|a_\bullet\right|\leq 1;
\eqf
if 
\eqi
\left|a_\bullet\right|=1,\lb{aextr}
\eqf 
the Kerr BH is said maximally rotating, or extremal. Despite mathematically possible, the actual physical existence of extremal BHs is debated; see, e.g., \citep{2025GReGr..57...60D,2025arXiv251201898C}, and references therein. 
Should it be $\left|a_\bullet\right|>1$, a naked singularity without a horizon would form \citep{1991PhRvL..66..994S}, that would be prohibited by the still unproven cosmic censorship conjecture \citep{1965PhRvL..14...57P,2002GReGr..34.1141P} in the case of the gravitational collapse of a material body; for recent discussions and reviews, see, e.g., \citep{1999JApA...20..233P,2020IJMPA..3530007O,2021FoPh...51...42L}, and references therein. Remarkably, the absence of a horizon would imply  the possibility of
causality violations because of closed timelike curves \citep{1978GReGr...9..155C,1979GReGr..10..335D}.

It should be noted that \rfr{ONH} is based on the averaged orbital precessions of the inclination and the longitude of the ascending node of a test particle induced by the Newtonian quadrupole mass moment and the 1pN gravitomagnetic field of a rotating material body \citep{2025MNRAS.537.1470I,2025PhRvD.111d4035I}. As such, in principle, it is not guaranteed that they also hold if the primary is a Kerr BH. Indeed, to date, no material source for the Kerr metric has yet been found, despite notable efforts \citep{1967JMP.....8.1477C,1967PhRv..159.1070H,1968PhRv..167.1180H,1968PhRv..172.1291W,1970PhRvD...2..641I,1978AnPhy.112...22K,1981NCimB..62..273H,1982JMP....23.2339H,1990NCimB.105..365H,1997CQGra..14.1883D,
2001NCimB.116.1009A,2005JPhCS...8...13P,2006IJMPD..15.1441V,2010IJMPD..19.1783V,2013IJMPD..2250051K,2014EPJC...74.2865A,
2017PhRvD..95b4003H,2025Univ...11...23C}. Thus, the latter only describes the exterior spacetime of a spinning BH.
Nonetheless, a direct calculation to the order of $\mathcal{O}\ton{1/c^2}$ of the geodesic equations for a test particle  moving in the Kerr metric, expressed in terms of the Kerr-Schild quasi-Cartesian coordinates \citep{1963PhRvL..11..237K,1965qssg.conf...99K}, shows that \rfr{ONH} holds also for a Kerr BH, characterized by \rfrs{SBH}{QBH}, provided that the formal replacement 
\eqi
J_2\rightarrow -\rp{Q_2}{MR_\mathrm{e}^2}\lb{J2}
\eqf 
is done. In \rfr{J2}, $J_2$ and $R_\mathrm{e}$ are the dimensionless quadrupole mass moment and equatorial radius, respectively, of the oblate material body acting as primary.

\begin{figure}[ht!]
\centering
\begin{tabular}{cc}
\includegraphics[width = 7.5 cm]{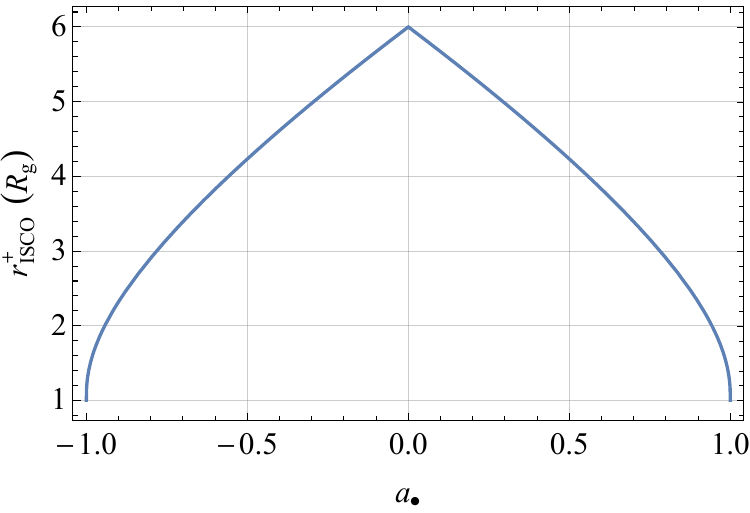}&\includegraphics[width = 7.5 cm]{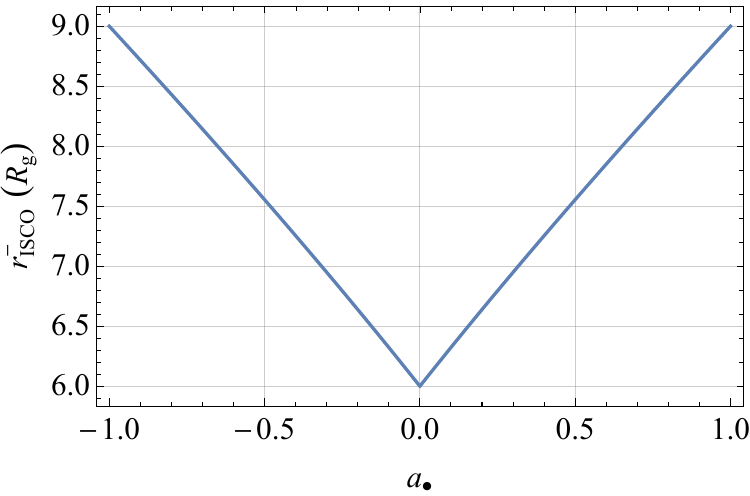} \\
\end{tabular}
\caption{Radii $r_\mathrm{ISCO}^{\pm}$ of the prograde ($+$, left panel) and retrograde ($-$, right panel) ISCOs, in units of gravitational radii $R_\mathrm{g}$, as functions of the dimensionless spin parameter $a_\bullet$ of a Kerr BH.}\label{Fig:1}
\end{figure}

In view of the following developments, it is useful to recall that, for a Kerr BH, the radius of the innermost stable circular orbit (ISCO)  \citep{1972ApJ...178..347B}
\eqi
\rp{r^+_\mathrm{ISCO}}{R_\mathrm{g}} = 3 + Z_2 - \sqrt{\ton{3 - Z_1} \ton{3 + Z_1 + 2 Z_2}}\lb{risco+} 
\eqf
in the BH's equatorial plane ranges from $6\,R_\mathrm{g}$ ($a_\bullet=0$) to $1\,R_\mathrm{g}$ ($\left|a_\bullet\right| = 1$) if $\boldsymbol{\hat{k}}_\bullet$ and $\boldsymbol{\hat{h}}$ are parallel (prograde motion), while its maximum value $9\,R_\mathrm{g}$ occurs from \citep{1972ApJ...178..347B}
\eqi
\rp{r^-_\mathrm{ISCO}}{R_\mathrm{g}} = 3 + Z_2 + \sqrt{\ton{3 - Z_1} \ton{3 + Z_1 + 2 Z_2}}\lb{risco-} 
\eqf
for $\left|a_\bullet\right| =1$ if $\boldsymbol{\hat{k}}_\bullet$ and $\boldsymbol{\hat{h}}$ are antiparallel (retrograde motion).
In \rfrs{risco+}{risco-}, it is \citep{1972ApJ...178..347B}
\begin{align}
Z_1 \lb{Z1} & := 1 + \ton{1 - a_\bullet^2}^{1/3}\qua{\ton{1 + a_\bullet}^{1/3} + \ton{1 - a_\bullet}^{1/3}}, \acap
Z_2 \lb{Z2} & := \sqrt{3 a_\bullet^2 + Z_1^2}.
\end{align}
If $a_\bullet=0$, \rfrs{risco+}{Z2} yield
\eqi
r^+_\mathrm{ISCO} = r^-_\mathrm{ISCO} = 6\,R_\mathrm{g}.
\eqf
Figure \ref{Fig:1} shows the plots of $r^{\pm}_\mathrm{ISCO}$ as functions of $a_\bullet$ according to \rfrs{risco+}{Z2}.
It turns out that the ISCOs' radii are independent of the sign of $a_\bullet$; moreover, it is $r^{-}_\mathrm{ISCO} > r^{+}_\mathrm{ISCO}$.
%
\section{Constraining the parameter space}\lb{Sec:3}
The strategy adopted here consists of identifying the phenomenologically measured precessional frequency of the disk-jet coprecession
\eqi
0.297\,\mathrm{d}^{-1}\leq \left|\Omega_\mathrm{obs}\right|\leq 0.347\,\mathrm{d}^{-1}\lb{cop}
\eqf 
with  \rfr{ONH}. The latter is written with \rfrs{SBH}{QBH} in order to have the NH precessional angular speed $\Omega_\mathrm{NH}$ for a Kerr BH. Then, it is considered as a function $\Omega_\mathrm{NH}\ton{x,y,z,\ldots}$ of some system's key physical and orbital parameters as independent variables $x,y,z,\ldots$. Finally, the condition that it must lie within the range of \rfr{cop} is imposed, namely
\eqi
\Omega_\mathrm{obs}^\mathrm{min}\leq\left|\Omega_\mathrm{NH}\ton{x,y,z,\ldots}\right|\leq\Omega^\mathrm{max}_\mathrm{obs}\lb{condiz}
\eqf 
is assumed. In this way, allowed regions in the parameter space $\grf{x,y,z,\ldots}$ can be obtained.   

The reason for which only the absolute value of the precessional frequency or, equivalently, of the precession period $T_\mathrm{prec}$ can be used resides in the fact that, in such kind of scenarios, one is not allowed to measure the full precession in the three-dimensional space, being only its projection on the plane of the sky accessible to direct observations. This fact can be understood more formally as follows. Let $\boldsymbol{\hat{\rho}}$ the unit vector of the line of sight: its angle with respect to $\boldsymbol{\hat{k}}_\bullet$ is $\theta_\mathrm{obs}$. Since the SMBH's spin stay fixed, $\theta_\mathrm{obs}$ is constant. The projection of the disk-jet coprecession onto the line of sight is
\eqi
\left|\dert{\boldsymbol{\hat{h}}}{t}\right|_\rho := \dert{\boldsymbol{\hat{h}}}{t}\boldsymbol\cdot\boldsymbol{\hat{\rho}}.\lb{proj} 
\eqf
According to \rfr{dhdt}, \rfr{proj} can be written as
\eqi
\left|\dert{\boldsymbol{\hat{h}}}{t}\right|_\rho = \Omega\boldsymbol{\hat{k}}_\bullet\boldsymbol\times\boldsymbol{\hat{h}}\boldsymbol\cdot\boldsymbol{\hat{\rho}}=
\Omega\boldsymbol{\hat{\rho}}\boldsymbol\times\boldsymbol{\hat{k}}_\bullet\boldsymbol\cdot\boldsymbol{\hat{h}}=\Omega\sin\theta_\mathrm{obs}\cos\upgamma\ton{t},
\lb{buga}\eqf
where $\upgamma\ton{t}$ is the time-dependent angle between the constant vector $\boldsymbol{\hat{\rho}}\boldsymbol\times\boldsymbol{\hat{k}}_\bullet$, whose modulus is $\sin\theta_\mathrm{obs}$, and the precessing unit vector $\boldsymbol{\hat{h}}$. Since $\sin\theta_\mathrm{obs} = \sin\ton{\pi-\theta_\mathrm{obs}}$, the latter corresponding to $-\boldsymbol{\hat{k}}_\bullet$, \rfr{buga} cannot discriminate between the two possible senses of the projected precession.

\subsection{The parameter space in the most general case}
\begin{figure}[ht!]
\centering
\begin{tabular}{cc}
\includegraphics[width = 7.5 cm]{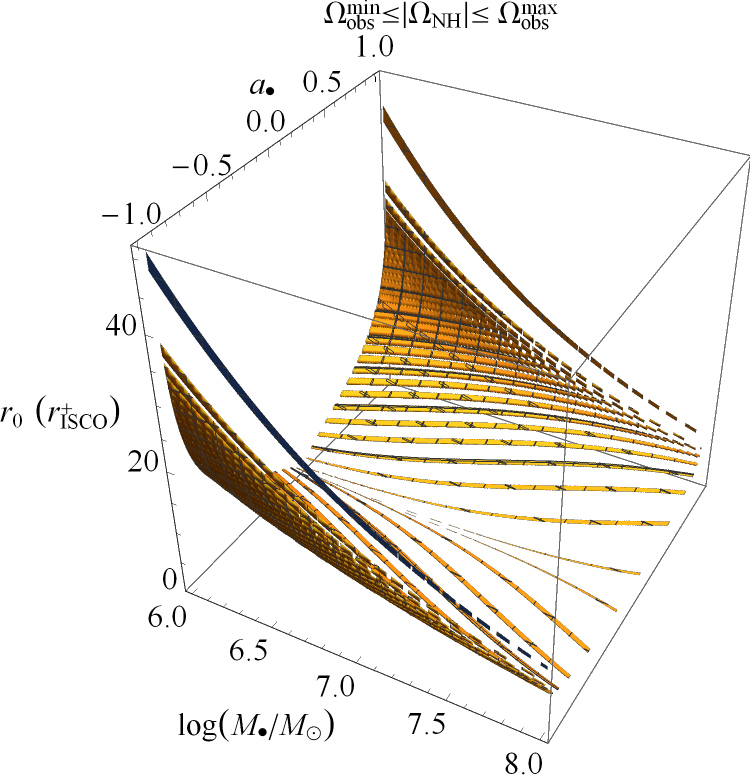}&\includegraphics[width = 7.5 cm]{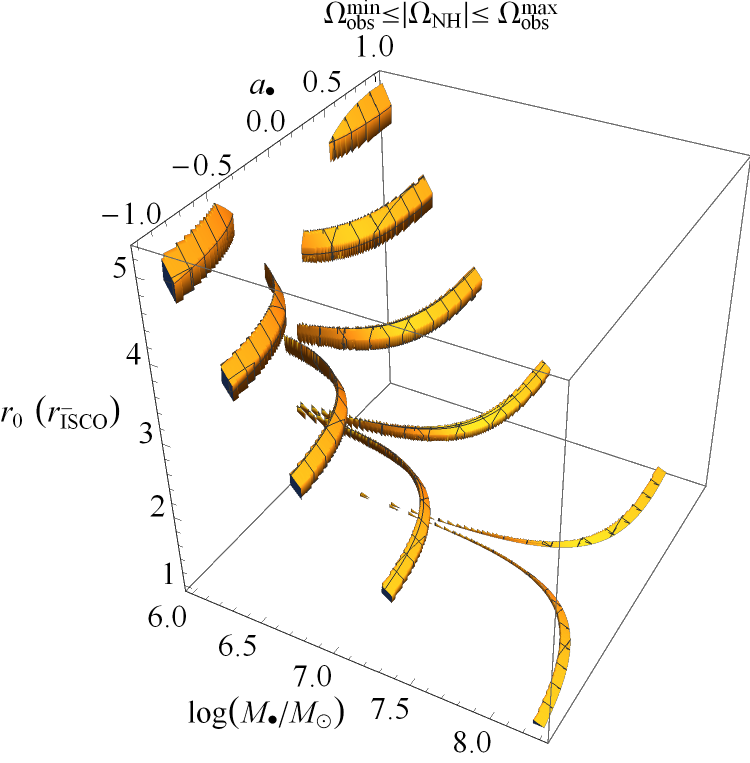} \\
\end{tabular}
\caption{Permitted regions in the three-dimensional parameter space $\grf{\log\ton{M_\bullet/M_\odot},a_\bullet,r_0}$ obtained by imposing the conditions that the absolute value of the NH precessional frequency of \rfr{ONH}, calculated with \rfrs{SBH}{QBH}, lies within the experimental range of \rfr{cop} for the observed disk-jet coprecessional one, and that the  radius $r_0$ of the effective test particle orbit is larger than that of the prograde ($r^+_\mathrm{ISCO}$, left panel) and retrograde ($r^-_\mathrm{ISCO}$, right panel) equatorial ISCOs. The best estimate of the colatitude $\theta_i$ of the orbital angular momentum is taken from Table \ref{Tab:1}, while $\log\ton{M_\bullet/M_\odot}$ and $a_\bullet$ are allowed to vary within their expected full ranges $6\lesssim \log\ton{M_\bullet/M_\odot}\lesssim 8$ \citep{1975Natur.254..295H,2023ApJ...959..117F} and $-1\leq a_\bullet \leq1$, respectively.}\label{Fig:2}
\end{figure}

By making the following broad assumptions 
\begin{enumerate}
\item The cause of the TDE is a SMBH, so that $6\lesssim\log\ton{M_\bullet/M_\odot}\lesssim 10$ \citep{2023ApJ...959..117F}
\item \lb{disr} For tidal disruption to actually occur outside a SMBH, accompanied by detectable flares, it must be $\log\ton{M_\bullet/M_\odot}\lesssim 8$ \citep{1975Natur.254..295H}
\item The external spacetime of the SMBH is described by the Kerr metric
\item The disk likely exhibits a not excessively large tilt to the SMBH's spin axis due to intervening non-gravitational effects of electromagnetic nature, so that $\boldsymbol{\hat{k}}_\bullet\boldsymbol\cdot\boldsymbol{\hat{h}}\lesssim 1$
\item The radius of the effective orbit should be at least larger than that of the SMBH's ISCO
\end{enumerate}
the condition of \rfr{condiz} preliminary yields the allowed regions in the three-dimensional parameter space $\grf{\log\ton{M_\bullet/M_\odot},a_\bullet,r_0}$ depicted in Figure \ref{Fig:2} for both prograde and retrograde ISCOs. About the point \ref{disr}, it can be explained as follows. If the tidal disruption radius 
\eqi
r_\mathrm{t} =\ton{\rp{M_\bullet}{M_\star}}^{1/3} R_\star\lb{rtidal}
\eqf
of a star with mass $M_\star$ and radius $R_\star$ is smaller than the SMBH's event horizon, the star is swallowed whole before being tidally torn apart, and no flares at all are produced. Indeed, a main sequence star moving along an elliptical orbit around a SMBH is dismembered if its pericenter falls within $r_\mathrm{t}$ \citep{1988Natur.333..523R,1989IAUS..136..543P,1989ApJ...346L..13E}. The aforementioned condition occurs just for $\log\ton{M_\bullet/M_\odot}\gtrsim 8$ \citep{1975Natur.254..295H}. 
In obtaining them, \rfrs{risco+}{Z2}, valid for equatorial orbits, were used.
Actually, the orbital plane of the effective orbit somehow representative of the precessing disk is not exactly aligned with the hole's equator; thus, strictly speaking, \rfrs{risco+}{Z2} are not valid. Nonetheless, they could be considered rather good approximations. Indeed, since the orbit's tilt to the hole's equator  is assumed to be somewhat small, the left panel of Figure 5 of \citet{2024ApJ...966..226K} shows that, for a turning point's angle, dubbed $\theta_\star$ as in \citet{2024ApJ...966..226K}, close to $90^\circ$, the radii of both the prograde and retrograde innermost stable spherical orbits (ISSOs), extending above and below the BH's equatorial plane for an inclined orbit, essentially agree with those for the ISCOs occurring in the equatorial case. An explanation of the aforementioned angle goes as follows. In standard spherical coordinates, adopted in the Boyer-Lindquist form of the Kerr metric \citep{1967JMP.....8..265B}, the coordinate $z_\mathrm{P}\ton{t}$ of a point P moving at constant distance $r_0$ from the center is $z_\mathrm{P}\ton{t} = r_0\cos \theta\ton{t}$, where $\theta\ton{t}$ is the colatitude angle reckoned from the reference $z$ axis which, in this case, is the SMBH's spin direction. In standard celestial mechanics, it is written as $z_\mathrm{P}\ton{t} = r_0\sin\theta_i\sin u\ton{t}$, where\footnote{In celestial mechanics, it is customarily denoted by $I$ \citep{2000ssd..book.....M}. Here, for the sake of clarity, it is preferred to use the same symbol $\theta_i$ adopted for it by \citet{TDELTCinesi2025}.} $\theta_i$ is the inclination of the orbital plane to the reference $\grf{x,y}$ plane which, in this case, is the SMBH's equator, and $u\ton{t}$, which is the argument of latitude, represents the particle's orbital phase. By equating both expressions, one has $\cos\theta\ton{t}=\sin\theta_i\sin u\ton{t}$. The turning points are intended to be those where $|z_\mathrm{P}|=|z_\mathrm{P}|^\mathrm{max}$: the corresponding colatitudes occur at $\theta_\star$ and $\pi - \theta_\star$. Thus, the angle of the first turning point is $\theta_\star=\arccos\ton{\sin\theta_i}$.
%

From Figure \ref{Fig:2}, it turns out that there are much more allowed regions for prograde orbits than for retrograde ones. In the first case, the effective radius may be as large as $50\,r_\mathrm{ISCO}^+$, while it can not be larger than $5\,r^-_\mathrm{ISCO}$ for the retrograde case. For the latter case, there are only about five allowed regions allowed corresponding to well detached multiples of $r^-_\mathrm{ISCO}$. 
Instead, the permitted domains are much more numerous for the prograde case. 
Such a feature, arising phenomenologically from the aforementioned quite general assumptions, is in agreement with the conclusion drawn by \citet{TDELTCinesi2025} with a more physical argument pertaining the measured disk's temperature and discussed in Section \ref{Sec:4}.
For prograde orbits, the largest radii correspond to the highest possible values of the spin parameters. The corresponding part of the allowed region is, however, rather small, making such a scenario highly disfavored with respect to the one characterized by smaller radii and moderate BH's rotation rates. On the other hand, it turns out that the minimum allowed value of $r_0$ amounts to $1\,r^{+}_\mathrm{ISCO}$ just for two extremely narrow regions corresponding to $a_\bullet\simeq\pm 0.1$. 
%
%
%
\subsection{Narrowing the parameter space with  the SMBH's mass}
A sensible narrowing of the allowed region in the parameter space can be obtained by using the observational constrain on the SMBH's mass, which was determined by \citet{TDELTCinesi2025}
with two independent methods irrespectively of any dynamical model for the disk-jet coprecession: the single-epoch virial method
and the M-sigma relation applied to spectral  measurements taken before the TDE. Both resulted in consistent estimates yielding the value reported in Table \ref{Tab:1}.
Furthermore, it contains also relevant parameters characterizing the space configuration of the system among which there is the angle $\theta_i$ entering $\boldsymbol{\hat{k}}_\bullet\boldsymbol\cdot\boldsymbol{\hat{h}}$ in \rfr{ONH}. From the Supplementary Material provided by \citet{TDELTCinesi2025}, it seems that also such an angle, along with the other ones listed in Table \ref{Tab:1} characterizing the orientation of the line of sight, was phenomenologically determined by fitting the $0.1-2$ keV X-ray lightcurve over a time span three months long (August-October 2024)  with a rigid precession model. Indeed, it appears that the hypothesis that the latter was explicitly due only to the LT effect  was explicitly used later to constrain $a_\bullet$ (see Eq. (S4) in \citet{TDELTCinesi2025}).

\begin{figure}[ht!]
\centering
\begin{tabular}{cc}
\includegraphics[width = 7.5 cm]{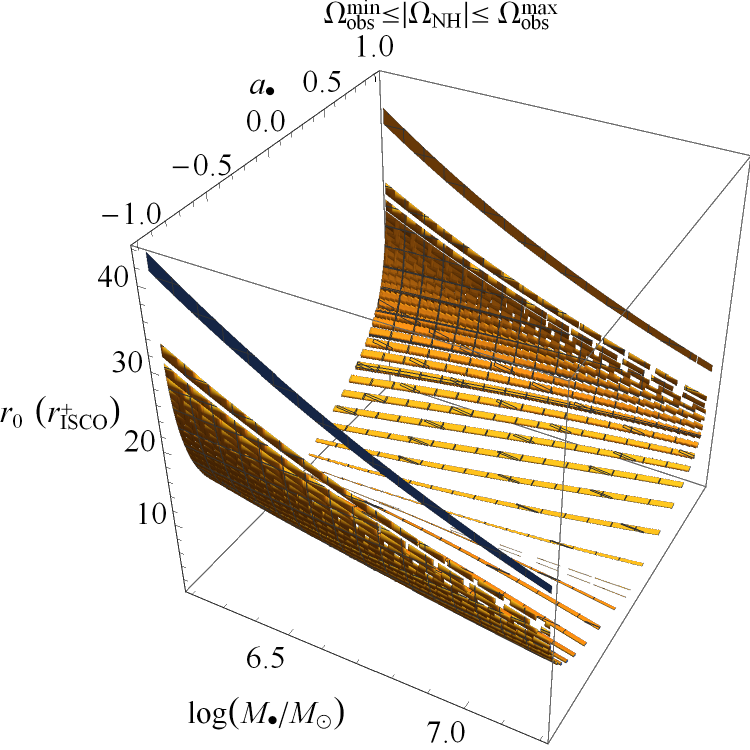}&\includegraphics[width = 7.5 cm]{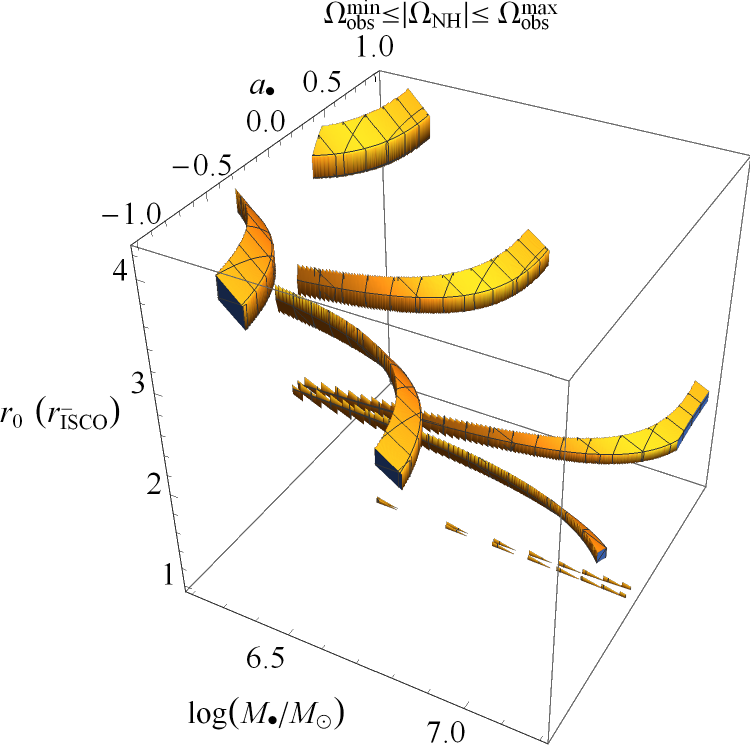} \\
\end{tabular}
\caption{Permitted regions in the three-dimensional parameter space $\grf{\log\ton{M_\bullet/M_\odot},a_\bullet,r_0}$ obtained by imposing the conditions that the absolute value of the NH precessional frequency of \rfr{ONH}, calculated with \rfrs{SBH}{QBH}, lies within the experimental range of \rfr{cop} for the observed disk-jet coprecessional one, and that the  radius $r_0$ of the effective test particle orbit is larger than that of the prograde ($r^+_\mathrm{ISCO}$, left panel) and retrograde ($r^-_\mathrm{ISCO}$, right panel) equatorial ISCOs. The range of variation for $\log\ton{M_\bullet/M_\odot}$ and the best estimate for the colatitude $\theta_i$ of the orbital angular momentum were taken from Table \ref{Tab:1}, while $a_\bullet$ is allowed to vary within its full range $-1\leq a_\bullet \leq1$.}\label{Fig:3}
\end{figure}

Figure \ref{Fig:3}, produced by using \rfr{ONH} calculated with the values of $\log\ton{M_\bullet/M_\odot}$ and $\theta_i$ reported in Table \ref{Tab:1}, depicts the new permitted region in the three-dimensional parameter space $\grf{\log\ton{M_\bullet/M_\odot},a_\bullet,r_0}$.

It should be noted that, for $\theta_i=14.5^\circ$, the first turning points' angle of the inclined orbit amounts to $\theta_\star=75.5^\circ=1.317\,\mathrm{rad}$. The left panel of Figure 5 of \citet{2024ApJ...966..226K} clearly shows that  the radii of the ISSOs are essentially identical to those of the ISCOs across the whole range for $a_\bullet$. Contrary to the prograde case, only three distinct allowed regions occur for a retrograde orbit corresponding to 2, 3 and $4\,r^-_\mathrm{ISCO}$. Thus, the prograde scenario is further strengthened  with respect to the retrograde one. It appears evident how high values of the spin parameter are quite unlikely, being admitted only for a quite restricted allowed sub-domain of the overall permitted domain. The largest possible value for $r_0$, corresponding to $a_\bullet =1$, is $40\,r^+_\mathrm{ISCO}$. 
Interestingly, it can be shown that, according to \rfr{rtidal}, the tidal disruption radius
of a sun-like star ranges from about $7.5$ to $35\,R_\mathrm{g}$ if it is calculated with the values of the Sun's mass and radius and the experimental range of values for $M_\bullet$ listed in Table \ref{Tab:1}, amounting to $16.1\ R_\mathrm{g}$ for $\log\ton{M_\bullet/M_\odot} = 6.7$. In \citep{TDELTCinesi2025}, the external radius of the physical disk was assumed to be twice the tidal disruption radius.
\subsection{Reducing the dimensionality of the parameter space by fixing the orbital radius}
In fact, by fixing the effective radius $r_0$ to some value, one is left with just two independent variables: $\log\ton{M_\bullet/M_\odot}$ and $a_\bullet$. 
To this aim, in the following, the reasonable, although bold assumption that 
\eqi
r_0 = r_\mathrm{t}
\eqf 
will be made.

\begin{figure}[ht!]
\centering
\begin{tabular}{cc}
\includegraphics[width = 15.0 cm]{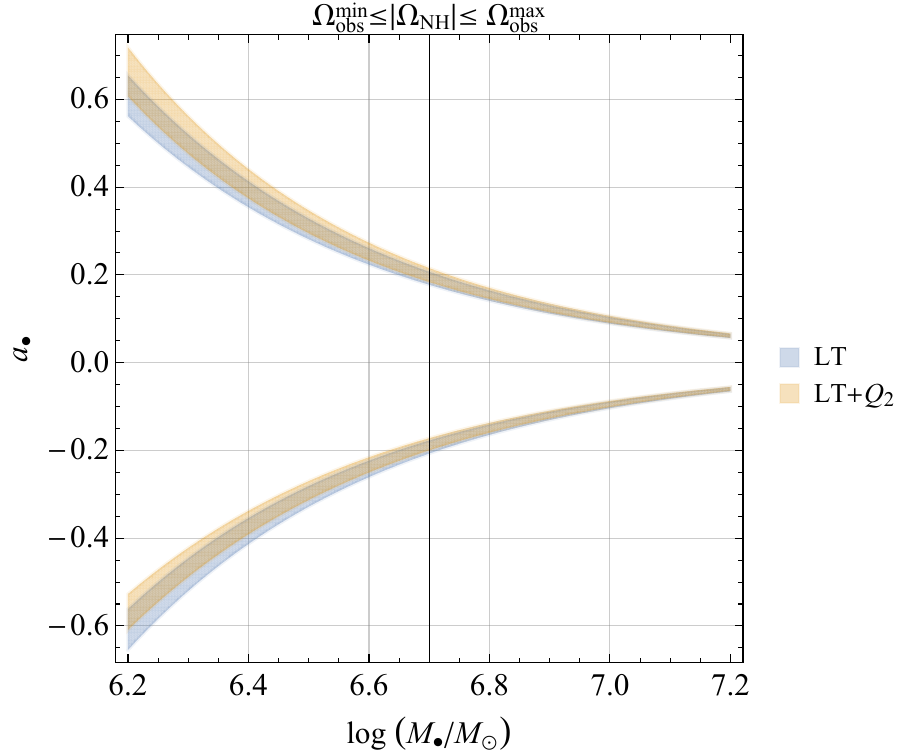}\\
\end{tabular}
\caption{Permitted regions in the two-dimensional parameter space $\grf{\log\ton{M_\bullet/M_\odot},a_\bullet}$ obtained by imposing the condition that the absolute value of the NH precessional frequency of \rfr{ONH}, calculated with \rfrs{SBH}{QBH}, lies within the experimental range of \rfr{cop} for the observed disk-jet coprecessional one. The radius $r_0$ of the effective test particle orbit is set equal to the tidal disruption radius of \rfr{rtidal}. The range of variation for $\log\ton{M_\bullet/M_\odot}$ and the best estimate of the colatitude $\theta_i$ of the orbital angular momentum were taken from Table \ref{Tab:1}, while $a_\bullet$ is allowed to vary within its full range $-1\leq a_\bullet \leq1$.}\label{Fig:4}
\end{figure}

Figure \ref{Fig:4} shows the allowed regions in the two-dimensional parameter space $\grf{\log\ton{M_\bullet/M_\odot},a_\bullet}$. It was obtained by using the values of the Sun's mass and radius in \rfr{rtidal}. Furthermore, it displays both the pure LT effect and the full NH bounds including also the impact of the SMBH's quadrupole $Q_2$, calculated with the value of $\theta_i$ reported in Table \ref{Tab:1}, in addition to the LT one. The resulting difference in the allowed ranges for $a_\bullet$ with and without $Q_2$ is of the order of about $0.01$.
It can be remarked that $r_\mathrm{t}=16.1\,R_\mathrm{g}$ is well larger than the maximum values of both the prograde and retrograde  ISCOs' radii. 

Now, the overall allowed range of values for the SMBH's spin parameter is 
\eqi
0.06\leq a_\bullet\leq 0.7,
\eqf
where the width of the permitted intervals of $a_\bullet$ for given values of $\log\ton{M_\bullet/M_\odot}$ ranges from about $0.01$ for $\log\ton{M_\bullet/M_\odot}\simeq 7.2$ to at most $0.1$ for $\log\ton{M_\bullet/M_\odot}\simeq 6.2$ .
Furthermore, values of the spin parameter larger than $\simeq 0.5$ look disfavored since they correspond to a very narrow range of observationally admitted values for the SMBH's mass. Instead, $82\%$ of the entire allowed region of Figure \ref{Fig:4} corresponds to $a_\bullet\lesssim 0.5$. In particular, $\log\ton{M_\bullet/M_\odot}=6.7$ gives 
\eqi
0.185\lesssim a_\bullet \lesssim 0.215,
\eqf
in substantial agreement with the range of positive values of \rfr{bound2} reported in \citet{TDELTCinesi2025} which, on the other hand, corresponds to about $47\%$ of the overall allowed region in Figure \ref{Fig:4}. Furthermore, as explained in Section \ref{Sec:4}, it is unclear if \rfr{bound2} can be really considered as a continuous range. 
\section{A discussion on the range for the BH's spin parameter by Wang et al. 2025}\lb{Sec:4}
In their Section 6.1, \citet{TDELTCinesi2025} adopted the following strategy to constrain the SMBH's spin parameter. As it will turn out, it relies upon  model-dependent assumptions.

First, they considered only the LT effect as cause of the observed disk-jet coprecession, without including also the SMBH's quadrupole. Accordingly, they wrote down their Equation (S4) modelling the precession period for a physical thick disk of finite extension as\footnote{In the original version of the Supplementary Materials of \citet{TDELTCinesi2025}, a typo occurred in Equation (S4) which displayed the reciprocal of the two multiplicative factors of \rfr{S4} containing $x_\mathrm{out}$ and $\zeta$.}
\eqi
T_\mathrm{prec} = \rp{\uppi GM_\bullet r^3_\mathrm{ms}}{a_\bullet c^3}\rp{x_\mathrm{out}^{5/2 - \zeta} - 1}{1 - x_\mathrm{out}^{-1/2 - \zeta}}\rp{\zeta + 1/2}{5/2 - \zeta}.\lb{S4}
\eqf
In \rfr{S4}, $r_\mathrm{ms}$ and $x_\mathrm{out}$ are dimensionless quantities denoting the radii of the ISCO and of the disk's outer boundary, respectively,  in units of gravitational radii. To this aim, the double of the tidal disruption radius as given by \rfr{rtidal} was taken by \citet{TDELTCinesi2025} for the disk's external radius; the internal radius was assumed by \citet{TDELTCinesi2025} equal to ISCO's. Finally,  $\zeta$ denotes\footnote{Here, the symbol $\zeta$ is adopted for it instead of $p$ used in \citet{TDELTCinesi2025} to avoid confusions with the semilatus rectum of the test particle's orbit.} a dimensionless power-law index for a disk's surface density profile for which the possible values $\zeta=0, 3/5, 3/4$ were adopted by \citet{TDELTCinesi2025}.

Then, \citet{TDELTCinesi2025} produced their \textbf{Fig. 6} (\textbf{A}), obtained as follows. For each given value of the parameter $\zeta$, they filled the space between the graphs of the absolute value of $T_\mathrm{prec}$, given by \rfr{S4} and calculated with the minimum and the maximum determined values of $\log\ton{M_\bullet/M_\odot}$, as a function of $a_\bullet$ getting three different gray-shaded regions. Importantly, they look different depending on whether $a_\bullet$ is positive or negative. Finally, \citet{TDELTCinesi2025} superimposed  blue horizontal lines corresponding to the experimental range for $T_\mathrm{prec}$ which intersect the previously obtained domains in two narrow zones of different widths, highlighted by pink vertical bands.
As a result, they claimed the following allowed continuous ranges for the SMBH's spin parameter 
\begin{align}
-0.46 \lb{bound1} & \leq a_\bullet \leq -0.14, \acap
0.11 \lb{bound2} & \leq a_\bullet \leq 0.35.
\end{align} 
The negative values were rejected by \citet{TDELTCinesi2025} since, according to them, they would lead to a larger ISCO and, consequently, a larger inner disk radius, in contrast with the measured hot temperature for it. 

\begin{figure}[ht!]
\centering
\begin{tabular}{cc}
\includegraphics[width = 7.5 cm]{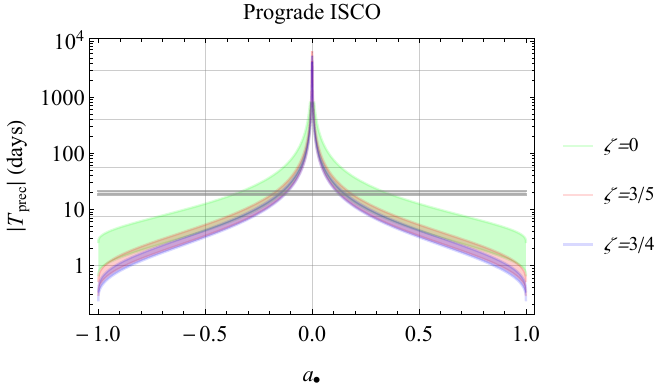}&\includegraphics[width = 7.5 cm]{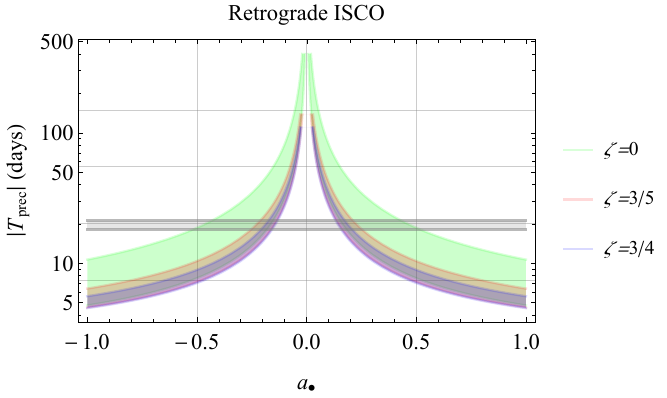}\\
\end{tabular}
\caption{Allowed regions in the $\grf{a_\bullet,T_\mathrm{prec}}$ plane obtained by plotting the absolute value of \rfr{S4} versus $a_\bullet$ for $\zeta=0,3/5,3/4$ \citep{TDELTCinesi2025}. Each shaded region is delimited by the log-linear plots obtained with the maximum and minimum values of $\log\ton{M_\bullet/M_\odot}$, as per Table \ref{Tab:1}. Left panel: prograde ISCO. Right panel: retrograde ISCO.}\label{Fig:5}
\end{figure}

Figure \ref{Fig:5} displays the allowed regions in the $\grf{a_\bullet,T_\mathrm{prec}}$ plane reproduced in the present study according to the absolute value of \rfr{S4}  for both prograde (left panel) and retrograde (right panel) ISCOs.
It turns out that the prograde and retrograde panels of Figure \ref{Fig:5} for $a_\bullet >0$ and $a_\bullet <0$, respectively, agree well with the half-planes $a_\bullet>0$ and $a_\bullet<0$  displayed in \textbf{Fig. 6} (\textbf{A}) of \citet{TDELTCinesi2025}, respectively. At this point, a possible misunderstanding may arise about the statement by \citet{TDELTCinesi2025} on the negative values of $a_\bullet$.  In fact, it is not the negative sign of $a_\bullet$ that makes the radius of the ISCO larger, but its retrograde character, as shown by \rfrs{risco+}{Z2} and Figure \ref{Fig:1}. 
Thus, the physical reasons given by \citet{TDELTCinesi2025} for seemingly rejecting the negative values of $a_\bullet$ actually serve to exclude the retrograde ISCO, of whose picture they decided to show in \textbf{Fig. 6} (\textbf{A}) only the part with $a_\bullet <0$. Thus, the bound of \rfr{bound1} is somehow misleading since it refers just to retrograde ISCOs, which are ruled out. Given that only prograde ISCOs are likely permitted, \rfr{bound1} should be replaced by the specular interval of \rfr{bound2}, namely $-0.35\leq a_\bullet \leq -0.11$. 

The values $\zeta = 0, 3/5, 3/4$ were not chosen by \citet{TDELTCinesi2025} for any particular reason, as they are simply figures widely used in the literature. It must remarked that the density profile of accretion disks is, actually, highly uncertain \citet{2022hxga.book....3L}.

An equivalent approach to the one just followed, conveying the same amount of information and corresponding to the one adopted in Section \ref{Sec:3}, would consist in plotting  the permitted domains in the parameter space $\grf{\log\ton{M_\bullet/M_\odot},a_\bullet}$ for each value of $\zeta$ by imposing that \rfr{S4} lies within the experimental range for $T_\mathrm{prec}$ as reported in Table \ref{Tab:1}. 
\begin{figure}[ht!]
\centering
\begin{tabular}{cc}
\includegraphics[width = 7.5 cm]{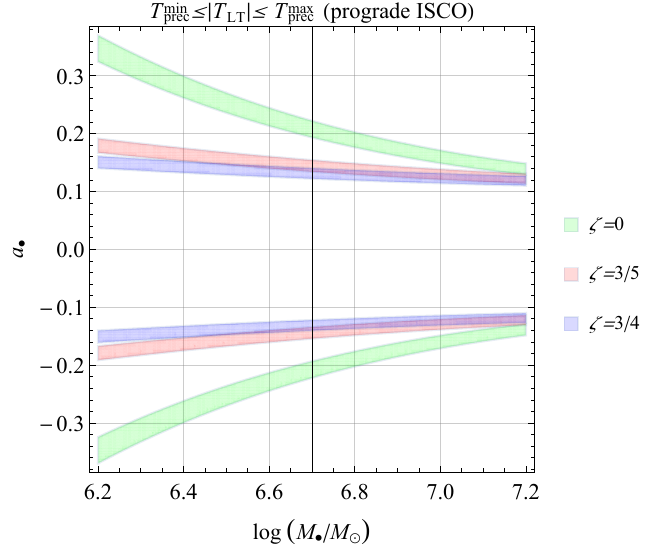}&\includegraphics[width = 7.5 cm]{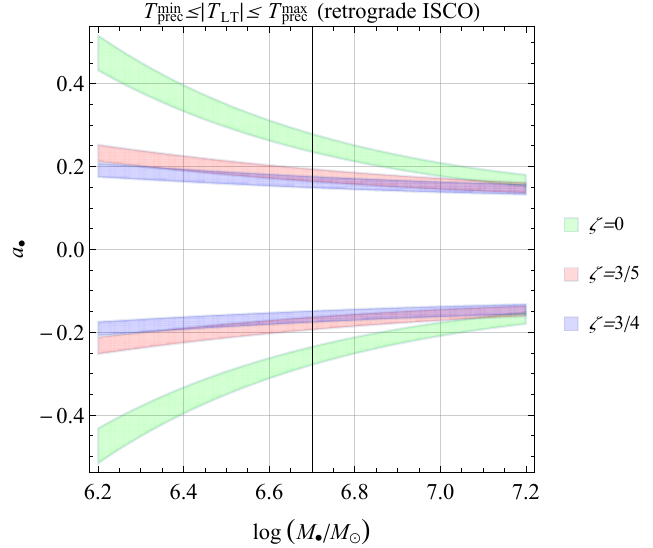}\\
\end{tabular}
\caption{Allowed regions in the $\grf{\log\ton{M_\bullet/M_\odot},a_\bullet}$ plane obtained for $\zeta=0,3/5,3/4$ \citep{TDELTCinesi2025} by imposing that the theoretical model of the absolute value of the period of the disk-jet coprecessions given by \rfr{S4} must lie within the experimental range of it as per Table \ref{Tab:1}. Left panel: prograde ISCO. Right panel: retrograde ISCO.}\label{Fig:6}
\end{figure}
Figure \ref{Fig:6}, implementing it by assuming $\left|T_\mathrm{prec}\right|$, shows that some interpretational issues may arise for the bound of \rfr{bound2}. Indeed,
both the panels of Figure \ref{Fig:6} seem to suggest that it should not be considered as a continuous range, being, instead, replaced by up to three distinct smaller subranges, one for each value of $\zeta$. It should be noted that only for the retrograde ISCO there is a very small domain centered around about $\log\ton{M_\bullet/M_\odot} \simeq 7.15-7.2,\,a_\bullet \simeq 0.1-0.13$ where all three regions overlap.

\section{\textcolor{black}{A critical examination of some of the assumptions on which this study is based}}\lb{Sec:5}
\textcolor{black}{The main assumption underlying this work is that the disk, which emits X-rays, and the jet, which emits radio waves, are strongly coupled, resulting in a coprecession with the observed period. In this regard, the following observations are in order.}
\textcolor{black}{In fact, the interpretation that the radio modulation arises from a coprecessing jet coupled to the X-ray emitting inner accretion raises, in principle, a timing issue that deserves to be addressed in further studies. If the radio emission originates at distances of order $10^{15}$ m from the event horizon, as usually inferred for synchrotron-emitting regions in TDE jets, the propagation time down the jet is approximately $3\times10^6$ s, corresponding to about 38 days. For mildly relativistic jet speeds, namely for $\qb_\mathrm{jet}:=v_\mathrm{jet}/c =0.85$, the propagation time remains of order $40-50$ days. Even allowing for relativistic travel-time effects that shorten the observed delay for small viewing angles, the expected lag between variability produced near the black hole traced by the X-rays and radio emission region could still be comparable to several days if not longer.}
\textcolor{black}{Now, in view of the fact that the reported quasi-periodicity in this source amounts to about 20 days, the characteristic propagation time to $10^{15}$ m is comparable to or larger than the entire oscillation period. In such a situation, one would generally expect the radio variability to be significantly phase-shifted relative to X-rays.}
\textcolor{black}{Such issues deserve further dedicated investigations.}
\textcolor{black}{Thus, this poses the problem how the observed radio periodicity remains closely correlated with the X-ray modulation under the above conditions. A possible explanation may be that the radio emission region lies substantially closer to the black hole than $10^{15}$ m. On the other hand,  relativistic projection effects may reduce the apparent lag.}

\textcolor{black}{Correlated with the above issue, there is also the fact that GRMHD simulations show jets that often align with the SMBH's spin and not necessarily with the outer disk, especially for magnetically arrested disks  (MAD).}

\textcolor{black}{Finally, in this paper, the disk-jet system is modelled as a single effective test particle orbit whose angular momentum vector precesses under the action of the Lense-Thirring effect. In fact, real TDE disks should experience differential gravitomagnetic torques, Bardeen–Petterson alignment, warping and internal viscosity. The disk is unlikely to precess rigidly unless $T_\mathrm{warp}\ll T_\mathrm{LT}$ which depends on poorly constrained parameters like, e.g., $\qa\tn$viscosity \citep{1973A&A....24..337S,2026arXiv260310997A} and the relative geometric thickness. Thus, rigorously speaking, it should be demonstrates that the TDE disk at hand  is in the rigid-precession regime.}

\textcolor{black}{All the above points well deserve further dedicated studies.}
\section{Summary and conclusions}\lb{Sec:6}
A simple 1pN analytical model of the LT precession of the orbital angular momentum of a fictitious test particle moving along an effective circular orbit around a rotating material object is able to explain the main features of the recently observed disk-jet coprecession in the TDE associated with the optical transient AT2020afhd driven by the spinning SMBH of almost ten million solar masses in the core of the galaxy LEDA 145386. 

A preliminary analysis based on quite general assumptions on TDEs shows that the allowed regions in the three-dimensional parameter space spanned by the SMBH's mass, its dimensionless spin parameter and the effective orbital radius,  are much more numerous for  prograde than for retrograde ISCOs, thus disfavoring the latter ones. Such permitted domains are obtained by only imposing the condition that the precession frequency analytically calculated with the aforementioned simple LT model lies within the experimental range for the measured rate of the disk-jet coprecession.

By using the measured interval of values of the SMBH's mass and setting the effective orbital radius equal to the tidal disruption one, narrower permitted domains in the two-dimensional mass-spin parameter space are obtained. While they are symmetric for negative and positive values of the spin parameter if only the LT effect is considered, the introduction of the further precession due to SMBH's quadrupole mass moment breaks this degeneracy. It turns out that, while the permitted range for the spin parameter extends from $0.06$ up to $0.7$, values larger than about $0.5$ are disfavored since they correspond to less than $20\%$ of the overall allowed domain. By adopting the best estimate for the SMBH's mass, the spin parameter is constrained within $0.185-0.215$, in agreement with the recently published bounds.

It is explicitly demonstrated that both the LT and quadrupole formulas used here, based on precessional models about a spinning material body acting as primary, coincide with those obtainable from the expansion to the 1pN order of the geodesic equations of motion of a test particle in the Kerr metric describing the exterior spacetime of a rotating BH in general relativity. 

The same strategy applied to the LT-only precession model used in the literature, which envisages an extended accretion disk of finite size endowed with a mass density profile parameterized in terms of a power-law index, yields to corresponding distinct allowed regions for different values of the latter which essentially do not overlap; they are overall comprised within the spin parameter's interval $0.11-0.35$.

The approach presented here, successfully used in previous papers also for the measured precession of the jet emanating from the surrounding of the SMBH M87$^\ast$, can be extended to precessional effects in other analogous scenarios allowing to predict their main features.

\textcolor{black}{Further dedicated studies will have to confirm some of the assumptions on which this study is based.}

\section*{Data availability}
No new data were generated or analysed in support of this research.
\section*{Conflict of interest statement}
I declare no conflicts of interest.
\section*{Funding}
This research received no external funding.
\section*{Acknowledgements}
I thank Yanan Wang, Wei-Hua Lei, Weikang Lin and Yuzhu Cui for constructive discussions on a number of topics. \textcolor{black}{I am also deeply grateful to three anonymous referees for their constructive comments which greatly improved the manuscript}.

\bibliography{Megabib}{}
\end{document}